\documentclass{article}
\usepackage{spconf,amsmath,graphicx}

\title{ON ADDRESSING PRACTICAL CHALLENGES FOR RNN-TRANSDUCER}
\name{Rui Zhao, Jian Xue, Jinyu Li, Wenning Wei, Lei He, Yifan Gong}
\address{Microsoft Speech and Language Group}
\begin{document}
%\ninept
%
\maketitle
\begin{abstract}
In this  paper,  several  works  are  proposed  to  address  practical challenges for deploying RNN Transducer (RNN-T) based speech  recognition  system.   These challenges  are  adapting  a well-trained RNN-T model to a new domain without collecting the audio data, obtaining time stamps and confidence scores at word level.  The first challenge is solved with a splicing data method which concatenates the speech segments extracted from the source domain data.  To get the time stamp, a phone prediction  branch  is  added  to  the  RNN-T  model  by  sharing  the encoder  for  the  purpose  of  force  alignment.   Finally, we  obtain word-level confidence scores by utilizing several types of features  calculated  during  decoding  and  from  confusion  network.  Evaluated with Microsoft production data, the splicing data adaptation method improves the baseline and adaptation with the text to speech method by 58.03\% and 15.25\% relative word error rate reduction, respectively.  The proposed time stamping method can get less than 50ms word timing difference from the ground truth alignment on average while maintaining the recognition accuracy of the RNN-T model. We also obtain high confidence annotation performance with limited computation cost.
%The works in this paper are to address practical challenges for deploying RNN Transducer (RNN-T) based speech recognition system, which include adapting a RNN-T model to a new domain, obtaining time stamp and confidence scores at word level.  The first challenge is solved with a splicing data method which concatenates the speech segments from the existing data.  The word timings are obtained by force alignment with a phone prediction model sharing the encoder with RNN-T. The word-level confidence scores are calculated with confusion network based on several features from decoding.  Experimental results showed the splicing data method improves the baseline and adaptation with the text to speech method by 58.03\% and 15.25\% relative word error rate reduction, respectively.  The word timing method could get less than 50ms word timing difference on average without accuracy loss. The confidence score method could get high confidence annotation performance with limited computation cost.
\end{abstract}
\begin{keywords}
end-to-end, adaptation, word timing, RNN-T, confidence
\end{keywords}
\section{Introduction}
\label{sec:intro}

Recently, all-neural end-to-end (E2E) models \cite{Graves-CTCFirst, Graves-RNNSeqTransduction, chan2016listen, attentionisallyouneed, he2019streaming} have become the trend in automatic speech recognition (ASR)  due to its simple training and decoding processes, as well as the similar or even better accuracy compared with traditional hybrid speech recognition systems \cite{watanabe2017hybrid, sainath2020streaming, Li2020Developing}. Commonly used E2E ASR models include Connectionist Temporal Classification (CTC) model \cite{Graves-CTCFirst,miao2015eesen,li2018advancing}, Attention-based Encoder-Decoder (AED) model  \cite{chan2016listen}, and transducer models such as recurrent neural network transducer (RNN-T) \cite{Graves-RNNSeqTransduction}  and transformer-transducer (T-T) \cite{TT}. The industry has more interest in transducer models including RNN-T and T-T because these models can be streaming in a natural way \cite{he2019streaming, TT, Li2019RNNT, battenberg2017exploring,chiu2018state, Li2020comparison, yeh2019transformer, xiechen}. However, there are lots of practical challenges that need to be addressed in order to deploy all these models. In this paper, we will address some of them by using RNN-T as a backbone model. All solutions can be applied to other transducer models such as T-T.

%Among these E2E models, RNN-T model is more popular for real application due to its easy support for streaming mode. Many works have proved that RNN-T model could beat the state of art Hybrid model on general task (\cite{Li2020Developing, streamingdevice}). But there are still some challenges in applying the RNN-T model to real applications. We focus on addressing three of these challenges in this paper. 

The first challenge is to adapt a RNN-T model to a new domain without collecting the speech data of the new domain. Because there is no separate language model in RNN-T, we couldn't easily adapt it with text only data as what we do for a hybrid ASR system. Instead, we need to collect paired speech-text data of the new domain which contains enough speaker and environment variation to adapt the RNN-T  model. Such collection usually costs a lot on both time and money, which forbids fast model  adaption \cite{belladaptation}. Therefore, how to get a large amount of paired speech-text data with a small cost is a practical problem for the domain adaptation of RNN-T. 

Some methods have been proposed to solve this problem. The most popular method is to synthesize speech from the new domain texts using the text to speech (TTS) technology \cite{Li2020Developing, eva2020adaptation, peyser2019improving,baskar2019self,murthy2018effect}. For example, \cite{Li2020Developing} uses a multi-speaker neural TTS system to generate speech data using the text-only data of the new domain to adapt the RNN-T model. %\cite{eva2020adaptation} uses the same TTS system to improve the key word spotting accuracy for the key word appears rarely in the training data for RNN-T model. 
\cite{peyser2019improving} improves a general ASR model’s performance on the numeric data domain with TTS generated numeric speech data. \cite{murthy2018effect} improves OOV detection by adding missing syllable sounds using a cross-lingual TTS system. Although no real speech data needs to be collected, the TTS-based method has its limitations: 1) the speaker variation in TTS generated data is limited compared with the real production data used for ASR model training, 2) The cost of training a multi-speaker neural TTS model and the generation of synthesized speech is large. There are also other methods proposed for the domain adaptation of E2E models \cite{zhongdomain,domainlow}. However, the speech data for the new domain is still needed in these methods.
%, since it needs sufficient speech training data for each speakers and the high performance computational resources (like GPU) for training and data generation. 

There is also a requirement that RNN-T can provide functions such as time stamp and confidence measure at word level. Getting word timings with RNN-T is challenging because it's not designed for this: no time alignment information is used during training, therefore it cannot generate reliable word start and end time for every recognized word. To solve this problem, \cite{wordtime} applies time constraints during training to improve the word timing accuracy. It uses an extra token ``word boundary'' to mark the word start time and the last word piece of the word to mark the word end time. This method increases the training cost because adding ``word boundary'' for each word increases the training target label length. Applying time constraints to RNN-T models also causes significant accuracy degradation \cite{wordtime}. Confidence annotation is also a challenging task for RNN-T models. Most applications require word level or utterance level confidence scores, while output units of RNN-T models are usually subword units such as word pieces. In \cite{conf_slt, conf_google}, word-level confidence scores are obtained through aggregating confidence scores of constituent word pieces and directly modeling last word piece using an end-of-word mask. Although these methods are very effective, the rich information included in all the word pieces instead of the last one for every word is lost to some extent.

In this paper, we detail our solutions to these practical challenges. To adapt RNN-T without collecting new speech data, a novel splicing data generation method is proposed by concatenating the sampled speech segments corresponding to underlying words of target texts into new utterances. The sampled speech segments are extracted from the existing training data randomly. 
%In this method, firstly, the audio segments of each word in the text are extracted from the existing training data randomly. Then, they are concatenated to form the audio of the text. 
It has the following advantages compared with TTS-based speech generation method : 1) the cost is almost zero since it is only based on the existing speech data, without the need of training any extra model and the cost-consuming TTS speech generation, 2) the constructed data could cover more speaker and acoustic environment variation, which makes the model more robust. 
%People may argue the audio data generated in this way is not fluent in the junctions of each word. This may not hurt RNN-T model since it makes recognition decision after processing a segment of speech instead of frame by frame. 
%Besides, when adapting model with such data, they are mixed with the real general training data and the lower layers of encoder network are fixed. These tricks help to prevent the model from over-fitting to the generated data. 
The proposed method is evaluated by adapting a general RNN-T model to a new domain with the spliced data, achieving 15.25\% relative word error rate reduction  over the model adapted with TTS generated data. %The proposed method will be called the splicing data generation method in the following part of the paper. 

To provide reliable word timings for RNN-T, we propose adding a context independent (CI) phone prediction branch on top of the encoder of the RNN-T model. 
%They could be trained jointly within the multitask learning (MTL) framework, or the CE model could be trained individually by fixing the encoder if a well trained RNN-T model is given. During the decoding, 
The word timings are calculated by aligning the recognition results from the RNN-T model using the phone probability of each frame from the CI phone prediction model in the second pass. Since the CI phone prediction model shares the encoder with RNN-T model and the phone level alignment is very cheap. The extra computational cost of this second-pass alignment is acceptable. The experiments proved that the proposed method could get less than 50ms word timing difference on average compared with the ground truth while maintaining the recognition accuracy of the RNN-T model.  

Confidence annotation is treated as a binary classification task in our work. We first obtain  confidence features at word piece level directly from decoding, then aggregate them to word level features. Compared to aggregating confidence scores \cite{conf_slt, conf_google}, aggregating features could retain much richer information, thus making the final word-level confidence scores more reliable. Together with the features calculated from confusion network generated with a N-best list \cite{confusion_network, cn_jian}, we train a two-layer feed forward neural network to classify each recognized word as ``correct'' or ``incorrect''. Experimental results showed that such a method could achieve high confidence annotation performance with limited computation cost. 

\section{rnn-t}
\label{sec:rnnt}

\begin{figure}[t]
  \centering
  \includegraphics[width=0.7\linewidth]{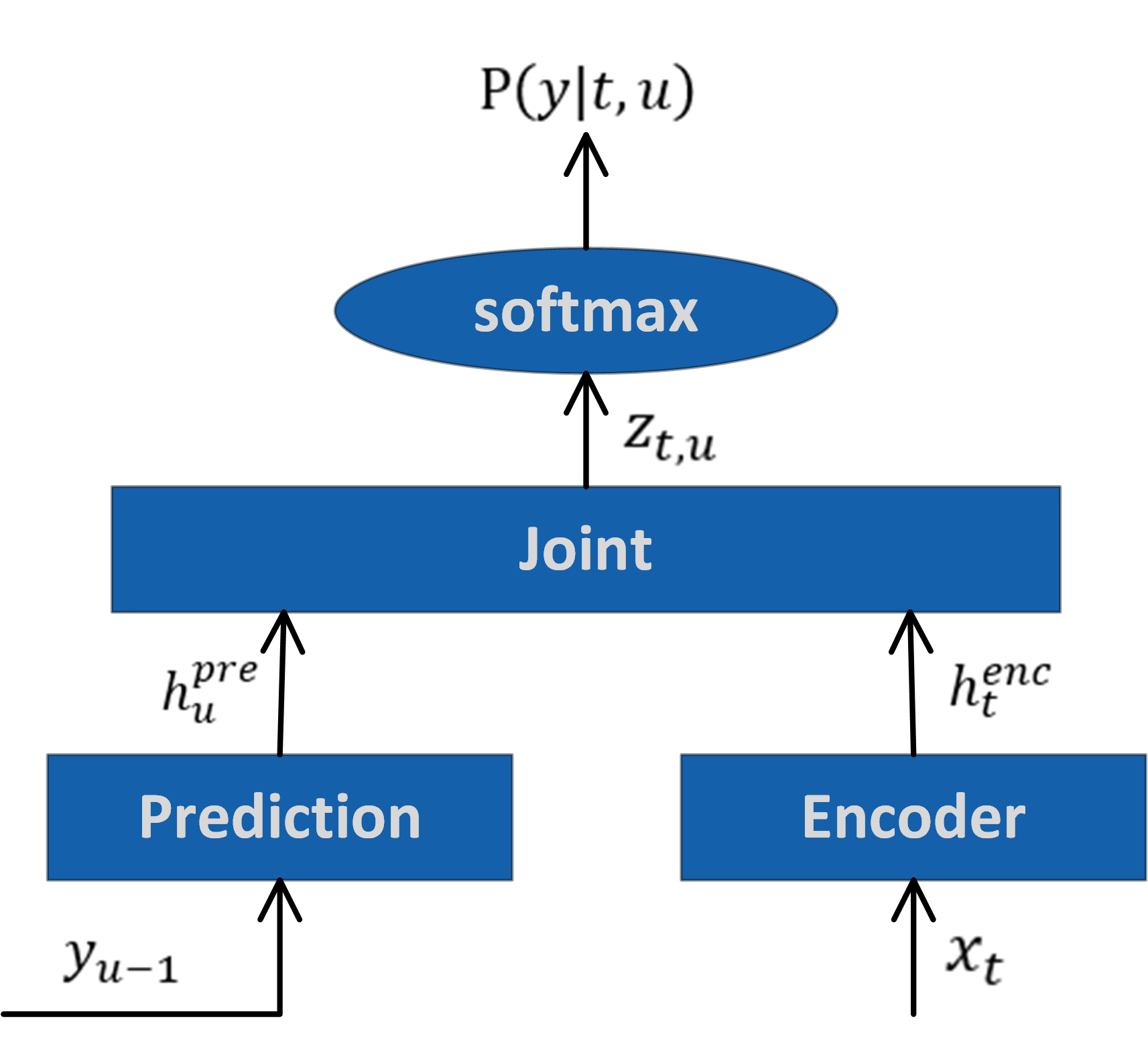}
  \caption{RNN-T model structure.}
  \label{fig:rnnt}
\end{figure}

A RNN-T model \cite{Graves-RNNSeqTransduction} consists of encoder, prediction, and joint networks as shown in Figure \ref{fig:rnnt}. The encoder network is analogous to the acoustic model in hybrid models, which converts the acoustic feature $x_t$ into a high-level representation $h_t^{enc}$, where $t$ is the time index. The prediction network works like a RNN language model, which produces a high-level representation $h_u^{pre}$ by conditioning on the previous non-blank target $y_{u-1}$ predicted by the RNN-T model, where $u$ is output label index.  The joint network combines the encoder network output $h_t^{enc}$ and the prediction network output $h_u^{pre}$ with a feed forward network as 
\begin{align}
z_{t,u}  &= f^{joint} (h_t^{enc}, h_u^{pre}) 
\end{align}

Then $z_{t,u}$ is connected to the output layer with a linear transform
\begin{align}
h_{t,u}=W_y z_{t,u} +b_y.\label{eq:h}
\end{align}
The final posterior for each output token $k$ is obtained after applying the softmax operation
\begin{align}
P(k|t,u)=softmax (h_{t,u})_k.
\label{eq:soft}
\end{align}

The loss function of RNN-T is the negative log posterior of output label sequence $\bf{y}$ given the input acoustic feature $\bf{x}$,
\begin{align}
L_{rnnt} = -ln P(\bf{y}|\bf{x}).
\label{eq:rnnt}
\end{align}

\section{Splicing data generation for domain adaptation}
\label{sec:splicing}

Domain adaptation is to adapt the well-trained source domain model to a target domain \cite{belladaptation} which usually has mismatched content from the source domain. Given the model was trained with a large amount of data, every word in the target domain usually has been observed while the word sequence is very different between the source and target domain. 

When there is no budget for collecting and labeling speech data in the new domain, we could use TTS to synthesize speech for given text as in \cite{Li2020Developing}. However, the speaker variation of the TTS generated audio is limited even with the state-of-the-art multi-speaker neural TTS model \cite{yanTTS} when compared with that in the training data for ASR models of real productions. Besides, the cost of training a multi-speaker neural TTS model and generating TTS speech with it is high due to two reasons: 1) It needs the real speech and the transcription corresponding to it from multi speakers to train the model. 2) TTS model training and speech data generation require high performance computational resources, such as GPU. 

\begin{figure}[t]
  \centering
  \includegraphics[width=1.0\linewidth]{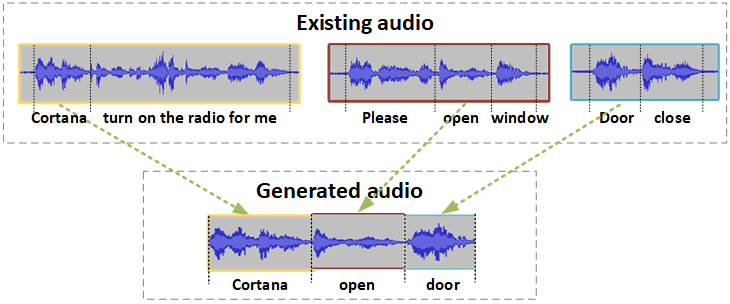}
  \caption{Spliced data generation.}
  \label{fig:splicing}
\end{figure}

In this paper, we propose a new method to generate speech data based on the existing ASR model training data. The proposed splicing data method generates new utterances by concatenating the audio segments of specific speech units (e.g.,  words) extracted from the existing training data. If there are multiple segments in the training data for the same word, we just randomly pick one segment.  Figure \ref{fig:splicing} shows the implementation of the proposed methods. The new audio for the given text ``Cortana open door'' is generated by concatenating the audio segments of words ``Cortana'', ``open'' and ``door''. These audio segments are extracted randomly from the existing training utterances. In this way, we can generate audio for almost any texts. If there is an out of vocabulary (OOV) word in the new domain, speech units corresponding to phones are used to construct speech for this OOV word.
It should be noted that there is no restriction when selecting audio segments, i.e., it is not required that the selected audio segments for one utterance should be from the same speaker, acoustic environment or corpus. They are all selected randomly. 
 
 The proposed method has clear advantages over the TTS-based audio generation method. Firstly, the cost is almost zero since it doesn't need any extra model or data. Secondly, the speech data generated with the proposed method is ``real'' speech at each segment, hence it has the potential to cover all the speakers and acoustic environments in the existing training data. Therefore the speaker and acoustic variation in such data is much higher than that in the TTS-generated speech data. The model trained with the data generated by the proposed method should be more robust than that trained with TTS data. %Lastly, the data generated by the proposed method contains different speakers and acoustic environment in one utterance, this will force the E2E model to learn more content information and ignore the speaker and acoustic environment changes. This will also help to improve the E2E model accuracy.  

People may argue that the generated audio in this way is not continuous at the transitions between words. As we know, one important feature of E2E models is that they make recognition decision after processing a segment of speech instead of frame by frame. For example, a RNN-T model will output ``blank'' if it isn't confident about what content the speech  until now should be. So this dis-fluency at the transitions between words won’t affect the RNN-T model too much. Besides, as shown in \cite{eva2020adaptation}, when adapting a RNN-T model with TTS generated data, the lower layers of the encoder should be fixed and the natural speech data should be mixed to combat the over-fitting to TTS data. The same trick is also used in the proposed method to reduce the side effects of the spliced data in the experiments in section \ref{adaptation} and it is proved to be effective by the results.  

\section{Word timing with phone prediction}
\label{sec:wordtiming}

To get accurate word timings of the recognition results with RNN-T model is challenging. Firstly, for a RNN-T model with word piece as the output token, it's difficult to get both the word start and end time. For example, when a frequent word is  toknenized into only one word piece, only the word start or end time could be estimated. Secondly, RNN-T model training is not guided by the ground truth alignment. Instead, all possible alignments are considered during the alignment, and the streaming model tends to delay its output to get better accuracy. To solve the above issues and get more accurate word timings, the authors in \cite{wordtime} add a ``word boundary'' token to get the word start time and use last word piece for word end time. Besides, the ground truth alignment is used to constraint the training. Although this method could help RNN-T model output better word timings, it results in recognition accuracy degradation due to the time constraint training. 

\begin{figure}[t]
  \centering
  \includegraphics[width=0.6\linewidth]{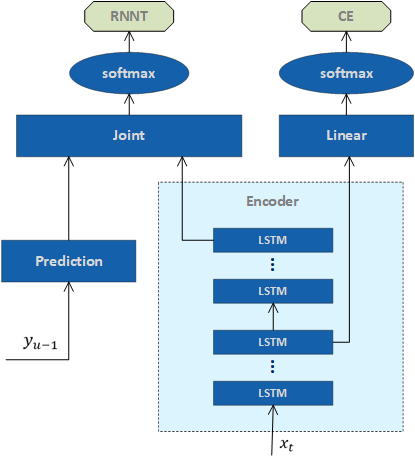}
  \caption{Word timing model structure}
  \label{fig:rnnt_ce}
\end{figure}

We propose  getting accurate word timings while maintaining the RNN-T model recognition accuracy by adding a CI phone prediction branch on top of the encoder. The CI phone prediction branch is used to obtain word timings with traditional force alignment method, instead of forcing RNN-T to do the work which it is not designed for. Specifically, the word sequence output from the RNN-T model is aligned with time based on the CI phone probability output in the second pass. The CI phone prediction model is trained with ground truth alignment using the frame wise cross entropy (CE) criterion. Therefore, the time alignment based on such model should be much more accurate than that obtained by a RNN-T model. 

The model structure of the proposed method is shown in Figure \ref{fig:rnnt_ce}. To save the computational cost, the CI phone prediction model shares the encoder with the RNN-T model.  In order to maintain a good recognition accuracy, it is better to not apply any time constraint to the RNN-T model during training. In contrast, the CI phone prediction model should be trained with ground truth alignment. If the whole encoder network is shared between these two models, they may have negative impacts on each other. Therefore, the CI phone prediction model shares only the lower layers of the encoder network of RNN-T model. %The experimental results also proved that sharing lower layers could obtain more accurate word boundary detection than sharing full encoder. 

The CI phone prediction and RNN-T models could be trained jointly with the multitask learning (MTL) method from scratch. In this case, the total loss is represented as follows by combining both RNN-T loss  in Eq. \ref{eq:rnnt} and CE loss. 
\begin{align}
L_{MTL} = \alpha L_{ce} + (1-\alpha)L_{rnnt}
\end{align}
Where $\alpha$ is the MTL weight. 

If we already have a well trained RNN-T model, we could also directly add the CI phone prediction branch and train it by fixing the encoder part. This ensures the RNN-T model won't be affected and saves training time. %Experimental results show the word timing output by this method is similar to that obtained by training CE and RNN-T jointly from the scratch. 

Because the CI phone prediction model shares the encoder with RNN-T model and the phone level alignment is fast. So the total cost of the second pass alignment should be quite small.

\section{Confidence annotation}
\label{sec:confidence}

We obtain word confidence scores using a binary classifier with features calculated from decoding and confusion network. The features calculated from decoding are at word piece level, and then we aggregate them to word-level features. We also generate confusion network features using different hypothesis scores. 

\subsection{Feature generation during decoding}
\label{sssec:subsubhead}
We generate three features for each word piece during decoding:
\begin{itemize}
    \item wp-prob: Log posterior probability of the word piece.
    \item hyp-prob: Log posterior probability of the partial hypothesis ended with the current word piece.
    \item neg-entropy: Negative entropy calculated over the vocabulary of the word pieces. 
\end{itemize}

Then we aggregate word piece level features into five word-level features:
\begin{itemize}
    \item avg-hyp-prob: Log posterior probability of the last word piece, averaged by the number of tokens in the partial hypothesis, including word pieces and ``blank'' token.
    \item min-wp-prob: Minimum of the log posterior probabilities for all word pieces in the word.
    \item avg-wp-prob: Average of the log posterior probabilities for all word pieces in the word.
    \item min-neg-entropy: Minimum of the negative entropy  for all word pieces in the word.
    \item avg-neg-entropy: Average of the negative entropy  for all word pieces in the word.
\end{itemize}

\subsection{Feature generation from confusion network}
%Confusion network was used in confidence annotation extensively before. During our work, 
We construct a word-level confusion network with the N-best list generated from decoding. We use two different scores for each hypothesis in the N-best list, one is the posterior probability of the hypothesis, another one is the length normalized posterior probability. Hence we obtain two confusion network based features, cn-prob and cn-norm-prob, for each word.

With the 7 word-level features described above, we train a 2-layer feed forward neural network to classify each word into two classes (correct and incorrect). The classifier is trained to minimize the binary entropy between the estimated confidence score $p$ and target class $c$, where $c$ is 1 if the word is correct, or 0 if the word is an insertion or substitution error. %We will investigate the effect of different features in the experiment section.
\begin{align}
L(c,p)  &= \sum_i(c_ilog(p_i)+(1-c_i)log(1-p_i)) 
\end{align}

\section{Experiments}
\label{sec:experiments}

The baseline RNN-T model was trained with 65 thousand (K) hours transcribed Microsoft data.%, which contains both the original data and the simulated data. 
%The simulated data is obtained by adding different noises and room impulse responses (RIR) to the original data to improve the robustness. 
All the data are anonymized with personally identifiable information removed.  The encoder network of the RNN-T model consists of 6 LSTM layers, with 768 nodes per layer. The prediction network consists of 2 LSTM layers, with 768 nodes per layer. The joint network is a single feed forward layer with 640 nodes. Singular value decomposition (SVD) \cite{xue2013restructuring} compression method is applied to further reduce the model size.  The feature is 80-dimension log Mel filter bank for every 10 milliseconds (ms) speech.  Three of them are stacked together to form a frame of 240-dimension input acoustic feature to the encoder network. The output targets are 4000 word piece units.

\begin{table}[t]
  \caption{WER for RNN-T models on a new domain }
  
  \label{tab:wer_adaptation}
  \centering
  \begin{tabular}{l|c}
        \hline
        \hline
			RNN-T	models			& 	WER(\%)       \\							
    	\hline
		baseline			& 9.27 \\
		adapted with TTS data 	&  4.59 \\ 
		adapted with spliced data  &  3.89 \\
    	\hline
    	\hline
  \end{tabular}
\end{table}

\begin{table*}[th]
  \small
  \caption{WER and word timing accuracy for different methods with RNN-T model}
  \label{tab:wordtiming}
  \centering
  \begin{tabular}{@{}l|c|c | c | c |c|@{}c}
    \toprule
    	&  \textbf{ RNN-T Baseline} & \textbf{RNN-T + CE }& \textbf{RNN-T + CE} & \textbf{RNN-T+CE}   & \textbf{ RNN-T + CE} &\textbf{RNN-T with  }   \\					
    	&   & \textbf{ update CE only}& \textbf{update CE only} & \textbf{MTL}   & \textbf{MTL} &  \textbf{time constraint } \\
    	\hline
    	\textbf{Shared encoder layers}& -  & \textbf{lower 4 }& \textbf{full } & \textbf{lower 4  }& \textbf{full } & - \\
    	 \hline
    	 WER(\%)  & 12.02  & 12.02 & 12.02 & 11.89 & 11.93 &12.84\\
    	 \hline
    	Ave. ST  $\Delta$ & - & 45ms &89ms  & 46ms & 86ms &120ms \\
    	\hline
    	Ave. ET  $\Delta$ & 202ms &42ms &85ms  & 44ms & 82ms & 97ms  \\
    	\hline
    	\% WS $<$ 200ms & - & 97.86 &94.42  & 97.69 & 94.19 &98.63 \\
    	\hline
    	\% WE $<$ 200ms & 49.88 & 97.80 &94.28 & 97.75 & 94.39 &98.62  \\
    	
   \bottomrule
  \end{tabular}
\end{table*}

\begin{table}[t]
  \caption{Confidence annotation results using each single feature and classifier }
  
  \label{tab:confidence}
  \centering
  \begin{tabular}{l | c | c }
        \hline
			\toprule			& 	AUPR-incorrect 		& 	AUPR-correct    \\							
    	\hline
		avg-hyp-prob			& 14.56			& 94.60 \\
    	\hline
		min-wp-prob			& 25.70			& 96.68 \\
    	\hline
		avg-wp-prob			& 21.71			& 96.45 \\
    	\hline
		min-neg-entropy			& 38.58			& 97.35 \\
    	\hline
		avg-neg-entropy			& 33.68			& 96.91 \\
    	\hline
		cn-prob			& 35.99			& 96.40 \\
    	\hline
    	\hline
		classifier			& 45.59			& 98.52 \\
    	\hline
    	\hline
		hybrid model \cite{li2020high}			& 23.86			& 97.28 \\
    	\hline    	
  \end{tabular}
\end{table}

\subsection{Domain adaptation}
\label{adaptation}

In this experiment, the baseline RNN-T model is adapted to a new command and control domain.%\footnote{This task is Microsoft production related, therefore no more details could be exposed}
 The testing data contains about 800 utterances collected in the real application environment. To adapt the RNN-T model into the new domain, first, the texts of the new domain are obtained by randomly parsing the grammar in the new domain and also using the crowd sourcing method, with about 200 thousand(K) sentences in total. Then, the speech of these texts are generated with either TTS or the proposed splicing data method. 

The multi speaker neural TTS model used to generate adaptation speech data is the best TTS model we have. It was trained by following the approach in \cite{yanTTS}. Firstly, a multi speaker neural TTS model is built with in-house TTS corpus, then it is fine-tuned by adding LibriSpeech  \cite{panayotov2015librispeech} data with thousands of speakers onboard. To evaluate the speech quality generated with this model, 10 speakers in LibriSpeech corpus are randomly selected, and 60 TTS samples are generated for each speaker. The subjective listening test (Mean Opinion Score, MOS) is carried out with crowd-sourcing judges. The results show the neural TTS quality is close to human recording of corresponding speakers. For the adaptation speech data generation, about 1000 speakers are selected, and for each speaker, 2.5K utterances are generated based on the texts randomly chosen from those 200K sentences. Then, the TTS speech data is doubled by adding simulated data, which is obtained by adding different noises and room impulse responses (RIR) to the original data to improve the robustness. In total, we got 5 million(M) utterances of TTS speech data for the new domain. 

In the splicing data method, for each text sentence, the audio segment for each word (or phone) in the text sentence is randomly extracted from the 65k hours general data. Similar to TTS data, a total of about 5M utterances are generated. 

When updating the baseline RNN-T model with the TTS or spliced data, we use two methods to prevent model from over fitting to the TTS or spliced data: 1) The lower 4 layers of the encoder network are frozen. 2) Similar amount of (about 5M utterances) normal speech data are randomly selected from the 65k hours data and mixed with the TTS or spliced data. %This recipe has been proved to be the optimal one when updating RNN-T model with TTS data in \cite{eva2020adaptation}. We didn't tune more about it in this experiment.  

%The audio data for these texts are generated with the proposed splicing data generation method as in \ref{fig:splicing}) and TTS method. After that, the general RNN-T model is adapted with splicing and TTS data separately .  

Table \ref{tab:wer_adaptation} shows the word error rate (WER) of the baseline    RNN-T model, as well as the RNN-T models adapted with the TTS and spliced data. Adapting with spliced data gets 58.03\% and 15.25\%  relative word error rate reduction over the baseline and the one adapted with TTS data, respectively. 

\subsection{Word timing}
\label{wordtiming}

%The RNN-T model structure in this experiment is the same as those in section 6.1 except the node number for each long short-term memory (LSTM) layer is 1024 instead of 768 for both prediction and encoder network. The training data is also the general 65 thousand hours data introduced in 5.1.  
In this experiment, the recognition accuracy evaluation set is a general test set which covers 13 application scenarios such as Cortana and far-field speech, containing a total of 1.8M words.The personally identifiable information is also removed for the testing data. 3500 utterances are randomly selected from the above 1.8M word set to evaluate the word timing accuracy. The reference word boundary is obtained by the force alignment with a traditional hybrid model. 

As in \cite{wordtime}, we measure the accuracy of word timings with below 4 metrics: 
\begin{itemize}
\item Average start time delta (Ave. ST $\Delta$): the average start time difference between the ground truth word start time and the estimated word  start time.
\item Average end time delta (Ave. ET $\Delta$): the average end time difference between the ground truth word end time and the estimated word end time.
\item Percentage of word start time less than 200ms (\% WS $<$ 200ms):  the percentage of word start time difference that is less than 200ms.
\item Percentage of word end time less than 200ms (\% WE $<$ 200ms):  the percentage of word end time difference that is less than 200ms.
\end{itemize}

Table \ref{tab:wordtiming} gives the WER and word timing accuracy for different methods. For the baseline model, the word end time is the emitting time of last word piece in this word. No word start time is estimated as explained in Section \ref{sec:wordtiming}. For the proposed methods, we examined the performance of two different training recipes: one is adding CI phone prediction model to a well trained RNN-T model and only updating the CI phone model by fixing the shared encoder part. Another is training RNN-T and CI phone  models with the MTL method from scratch. In this recipe, $\alpha$ is set to 0.1. We also build a RNN-T model based on the time constraint method in \cite{wordtime}. 

From the results, we can see all models trained with the proposed methods get smaller average word start/end time delta than the model trained with the time constraint method in \cite{wordtime}. The WER of RNN-T models trained with our proposed methods is not increased compared to the baseline RNN-T model, while the RNN-T model trained with time constraint method \cite{wordtime} has a higher WER. MTL Training from scratch gives a little lower WER than training only the CI phone prediction branch with CE.
These two different training recipes could get similar word boundary accuracy.  

We also compared the performance of sharing different number of encoder network layers: lower 4 layers or full encoder (6 layers).  Sharing lower 4 layers yields much better word boundary accuracy than sharing the full encoder: the average word timing difference is decreased from more than 80ms to less than 50ms. %This verifies our hypothesis in section \ref{ssec:timing} . 

The average decoding time is increased by about 1\% with the second pass alignment for word timings. 

\subsection{Confidence annotation}

%We use the same RNN-T model in section 6.2 to evaluate confidence annotation work. 
We conducted confidence experiments on the general test set described in section \ref{wordtiming}. %The classifier was trained with 95\% of the set, and we used residual 5\% of it to measure the performance.  WER (exclude deletion error) on the test set is 7.61. 
We use the area under precision-recall curve (AUPR) as the evaluation metric. We calculate AUPR for both correctly and incorrectly recognized words. The larger AUPR is, the better the confidence classifier performs. Table \ref{tab:confidence} summarizes the performance. 
From the results we can see the min-neg-entropy feature gives the best performance among all features. We further improve the performance by using the  classifier to combine all features together. In the last row, we also listed the confidence performance obtained with the method in \cite{huang2013predicting} of the hybrid model \cite{li2020high} as a reference.

We also evaluated the computation cost caused by the confidence annotation function during online decoding. For each utterance in the test set, we calculated the percentage of increased computation time after adding confidence annotation, compared to decoding the utterance without generating confidence features and score. The average percentage of increased computation time is 2.9\% for the whole test.
%, and Table \ref{tab:conf_cost} also gives the values for different percentiles. From the results we can see that the proposed confidence work doesn't bring much computation cost. Half of the utterances have less than 4.1\% increased computation time, and 90\% of the utterances have less than 8.3\% increased computation time. 

% \begin{table}[t]
%   \caption{Computation cost with confidence annotation}
%   \label{tab:conf_cost}
%   \centering
%   \begin{tabular}{l | c | c | c | c | c }
%   \hline
%   Percentile & 50\%  &  60\%  & 70\% & 80\%  & 90\%  \\
%   \hline
%   \hline
%   Increased time & 4.1\% &  4.7\%  & 5.2\%  & 6.1\%  & 8.3\%     \\
%   \hline
%   \end{tabular}
% \end{table}

\section{Conclusions}
\label{sec:illust}

In this paper, three challenges for deploying  RNN-T models to real applications have been addressed. To adapt a well-trained RNN-T model to a new domain without collecting new speech data, for all the  words in the text sentences of the new domain, we randomly extracted the corresponding audio segments from the source training data, and then concatenated them to form new speech utterances. Experimental result showed the model adapted with such data obtained 58.03\% and 15.25\% relative word error rate reduction over the baseline model and the model adapted with TTS generated data, respectively. To get accurate word timings of the recognition results, we added a context independent phone prediction branch by sharing the lower layers of the RNN-T encoder network to perform force alignment based on the phone probability of the phone prediction model. Experimental results showed that we could get less than 50ms average word boundary differences compared with the ground truth alignment without degrading the recognition accuracy of the RNN-T model. With the encoder sharing, the additional cost of the alignment is about 1\%. To obtain reliable word-level confidence scores for the RNN-T model with word piece units, we designed 7 word-level confidence features and a binary classifier. Experimental results showed the effectiveness of all the features and the binary classifier can further boost performance. The additional computational cost increase of confidence measure function is about 2.9\%.

\section{Acknowledgement}
We thank Kshitiz Kumar at Microsoft for providing the confidence measure evaluation of the hybrid model. We also thank Yuhui Wang and Min Hu at Microsoft for the runtime support of the time alignment and confidence measure of the RNN-T model. 

\newpage
\bibliographystyle{IEEEbib}
\bibliography{strings,refs}

\end{document}